\documentclass[twocolumn,showpacs,english,prl,superscriptaddress,preprintnumbers,amsmath,amssymb,floatfix]{revtex4-1}

\usepackage[T1]{fontenc}
\usepackage[latin9]{inputenc}
\usepackage{dcolumn}
\usepackage{bm}
\usepackage{graphicx}
\usepackage{color}
\usepackage{esint}
\usepackage{babel}
\usepackage{amsfonts}
\usepackage{slashed}
\usepackage{enumerate}

\begin{document}

\title{Towards  realistic implementations of a Majorana surface code}

\author{L.A.~Landau}
\affiliation{Raymond and Beverly Sackler School of Physics and Astronomy,
Tel-Aviv University, Tel Aviv 69978, Israel}

\author{S.~Plugge}
\affiliation{Institut f\"ur Theoretische Physik,
Heinrich-Heine-Universit\"at, D-40225  D\"usseldorf, Germany}

\author{E.~Sela}
\affiliation{Raymond and Beverly Sackler School of Physics and Astronomy,
Tel-Aviv University, Tel Aviv 69978, Israel}

\author{A.~Altland}
\affiliation{Institut f\"ur Theoretische Physik,
Universit\"at zu K\"oln, Z\"ulpicher Str.~77, D-50937 K\"oln, Germany}

\author{S.M.~Albrecht}
\affiliation{Center for Quantum Devices and Station Q-Copenhagen,  
Niels Bohr Institute, University of Copenhagen, 
Universitetsparken 5, DK-2100 Copenhagen, Denmark}

\author{R.~Egger}
\affiliation{Institut f\"ur Theoretische Physik,
Heinrich-Heine-Universit\"at, D-40225  D\"usseldorf, Germany}
\date{\today}

\begin{abstract}
Surface codes have emerged as promising candidates for quantum information processing. Building on the previous idea to realize the physical qubits of such
systems in terms of Majorana bound states supported by topological semiconductor nanowires, we show that the basic code operations, namely projective stabilizer
measurements and qubit manipulations, can be implemented by conventional tunnel conductance probes and  charge pumping via single-electron transistors, respectively.
The simplicity of the access scheme suggests that a functional code might be in close experimental reach.
\end{abstract}


\maketitle

\textit{Introduction.---}In recent years, surface codes have established themselves as a potent platform for universal quantum information processing. The basic idea of a surface code (when operated as a so-called stabilizer code) is to implement few `logical' qubits -- the actual carriers of quantum information  -- via the
correlation of a large number of physical qubits \cite{Bravyi2001,Gottesman1997,Freedman2001,Raussendorf2007,Fowler2012,Terhal2015,Wen2015}. 
What at first sight appears to be a redundant scheme offers a number of powerful advantages: (i) a degree of tolerance to  errors orders of magnitude higher than that
of other approaches, (ii) the possibility to implement a code as a comparatively simple two-dimensional (2D) layout of  cells coupled by nearest-neighbor
interactions, and (iii) the fact that essential code operations, including error tracking without need of active error correction in Clifford operations and/or memory access, 
are controlled by classical software.  (While non-local stabilizer codes may allow for even higher error thresholds \cite{Terhal2015}, we here focus on local surface codes.)  
These are highly attractive features which, accordingly, come at a hefty price tag: a large number of physical qubits is required even for modest operations. 
In particular, logical non-Clifford gates (e.g., the T gate) are required for universality, whose fault-tolerant implementation would require magic state distillation 
\cite{Terhal2015,Bravyi2005}. Under these conditions, Ref.~\cite{Fowler2012} estimates that about $10^{3-4}$ physical qubits
are needed to encode a reasonably fault-tolerant information qubit, implying that about $\mathcal{O}(10^8)$ physical qubits are needed to run, say, serious
factorization algorithms for integers with $\mathcal{O}(10^2)$ decimals. Achieving maximal simplicity in the implementation and in the \textit{access} of individual
qubits will therefore be a decisive factor in advancing from $\mathcal{O}(1)$ qubits to functional systems. In this Letter we argue that semiconductor Majorana hardware
layouts offer striking and so far unnoticed advantages in this regard.

At this point, two major platforms for the realization of surface codes are under discussion, Josephson junction arrays, and Majorana bound state (MBS) networks,
respectively. The Josephson junction architecture builds on physical qubits that are an experimental reality, and impressive progress towards the generalization to qubit
assemblies has been made recently \cite{Barends2014,Jeffrey2014,Martinis2015}. By contrast, not even the building blocks of an MBS qubit have been implemented so far.
However, few years after the prediction~\cite{Alicea2012,Leijnse2012,Beenakker2013,Alicea2011} of semiconductor MBSs, and the observation of first experimental
signatures~\cite{Kouwenhoven2012}, a now available second generation of topological nanowires features robust proximity coupling to an adjacent
superconductor~\cite{Chang2015} and sharply defined proximity gaps~\cite{Krogstrup2015,Higginbotham2015}. It stands to reason that this sets the
stage not only to the clean isolation of midgap MBSs, but also to their coupling by tunneling bridges~\cite{Albrecht2015}, which will be instrumental to the definition
of physical qubits based on MBS ring exchange~\cite{Terhal2012}.

A principal advantage of the semiconductor platform is that it can be accessed in terms of \emph{single-step} protocols which do not require ancilla
qubits~\cite{VijayFu2015}. (This is to be compared to a five-step projective measurement requiring microwave resonators and one ancilla qubit per physical qubit 
in approaches based on bosonic implementations of quantum storage \cite{Fowler2012,Barends2014,Jeffrey2014,Martinis2015}.) However, the concrete Majorana code readout
procedures suggested so far rely on a combination of gate electrode operations, microwave irradiation, and subsequent measurement of the phase shifts in the
transmitted photon field in a topological insulator platform~\cite{VijayFu2015}, or on magnetic interferometry on the
single qubit level within a nanowire setting~\cite{Terhal2012}. Both schemes involve extensive hardware overheads which raises questions regarding scalability.

In this Letter, we show that MBS codes can be accessed in much simpler terms via ordinary tunnel electrodes and similarly established  components of 
semiconductor quantum electronics that can be integrated into the 2D code structure itself. Within this framework, the system is operated 
by tunnel conductance measurements and  single-electron transistor (SET) manipulations in combination with classical software protocols. 
Importantly, no radiation fields nor probing magnetic fluxes are required.

\begin{figure}[t]
  \centering
\includegraphics[width=8.5cm]{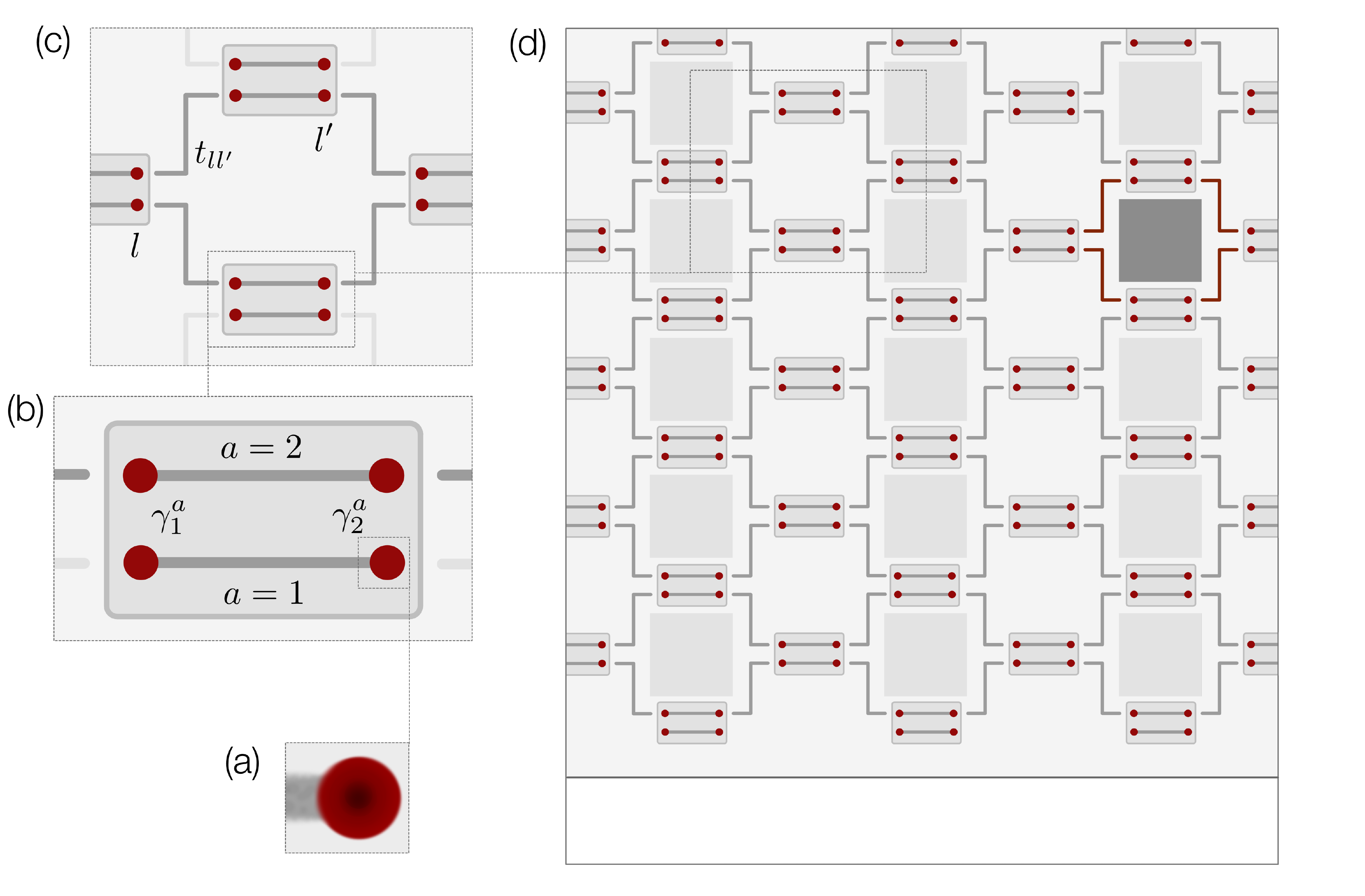}
  \caption{Structure of a surface code built on MBSs [red dot in panel (a)], few of which define a Majorana-Cooper
box [(b)]. Ring exchange of MBSs via tunneling bridges [(c)] defines a physical qubit through the eigenvalues $\pm 1$ of 
the plaquette operator ${\cal O}_n$, i.e., the product of all eight Majorana operators in the loop.
An assembly of qubits [(d)] can encode information qubits, e.g., the unit indicated by a dark shaded area.
For further discussion, see text.  }
  \label{f1}
\end{figure}

\textit{Semiconductor-based Majorana surface code.---} To start with, let us consider the schematic blueprint of a  Majorana code
depicted in Fig.~\ref{f1}. Its elementary building blocks are MBSs forming at the terminal points of spin-orbit coupled
(e.g., InAs) nanowires in a magnetic Zeeman field and proximitized by an $s$-wave superconductor
\cite{Alicea2012,Leijnse2012,Beenakker2013,Kouwenhoven2012,Chang2015,Krogstrup2015,Higginbotham2015}, 
cf.~the red dot in Fig.~\ref{f1}(a), where the wire and the superconductor are indicated by dark and light gray, respectively. We consider
sufficiently long wires such that the MBSs, which are described by anticommuting
operators $\gamma_{1,2}^{}=\gamma_{1,2}^\dagger$ with $\gamma_{j}^2=1$, represent
zero-energy states.  Pairs of wires ($a=1,2$) are contacted to the same floating mesoscopic superconducting island, see Fig.~\ref{f1}(b), such that the
entire `Majorana-Cooper box' \cite{Fu2010,BeriCooper2012,AltlandEgger2013,Beri2013,Plugge2015}
has a finite capacitance.  Each box then has a charging Hamiltonian $H_C=E_C ({\cal N}-n_g)^2$,
where ${\cal N}$ is the total number of electrons of the island relative to a backgate parameter $n_g$, and $E_C$ the 
single-electron charging energy. Assuming that $E_C$ defines the largest energy scale in the problem
and that $n_g$ is close to an integer value, $H_C$ will enforce charge quantization on the box. The resulting parity constraint, 
$\prod_{a=1,2} i \gamma_1^a \gamma_2^a= \pm 1$, reduces the four-fold ground-state degeneracy associated with four MBSs to 
a two-fold degeneracy. This effective `spin-$1/2$' degree of freedom is nonlocally encoded by the MBSs and
implies a `topological Kondo effect' \cite{BeriCooper2012} when a single box is contacted by normal leads. 
Instead, we here consider a 2D array of Majorana-Cooper boxes whose MBSs are tunnel-coupled to semiconductor wires connecting 
neighboring boxes $l$ and $l'$, see Fig.~\ref{f1}(c). Assuming a near instantaneous transmission through the (short)
tunneling bridges, we model the latter by weak static amplitudes $t_{ll'}$ with $|t_{ll'}|\ll E_C$.  The corresponding tunneling Hamiltonian reads 
$H_t=-\frac{1}{2} t_{l l'} \gamma_l \gamma_{l'} e^{i (\varphi_l-\varphi_{l'})/2}+{\rm h.c.}$, where   $\varphi_l$ is the fluctuating 
superconducting phase of  box $l$~\cite{BeriCooper2012,AltlandEgger2013,Beri2013}.

Under these assumptions, the low-energy excitations of the 2D array correspond to minimal-loop structures involving the product 
${\cal O}_n= \prod_{j=1}^8 \gamma_j^{(n)}$ of the eight MBSs,  $\gamma_j^{(n)}$, surrounding  loop no.~$n$ of the network. 
The plaquette operators $\mathcal{O}_n$ are the so-called stabilizers of the system. Their defining feature is that
they mutually commute (two stabilizers share either zero or two MBSs with each other) and possess only two eigenvalues,
$\mathrm{spec}(\mathcal{O}_n)=\pm 1$. This makes the stabilizers a primitive system of  qubits, the backbone of the surface code.
The corresponding effective low energy Hamiltonian 
\begin{equation}\label{sysham}
H_0 = - \sum_n  {\rm Re}(c_n) {\cal O}_n,\quad c_n=  \frac{5}{16 E_C^3} \prod t_{l_nl_n'},
\end{equation}
contains the products of the four tunnel amplitudes connecting the islands $l_n$ along the $n$th loop, where the  numerical prefactor follows from a Schrieffer-Wolff
transformation \cite{BravyiLoss2011}, see also Refs.~\cite{XuFu2010,Nussinov2012,Franz2015}. The projection to the lowest charge
sector in each box effectively removes any dependence on the phases $\varphi_l$ in Eq.~\eqref{sysham}. Notably, the code works even though the coefficients Re$(c_n)$ in
Eq.~\eqref{sysham} are uncorrelated random energies in general \cite{foot1}.

\textit{Operation principles.---}Let us recapitulate the essential steps by which a local surface code is operated~\cite{Bravyi2001,Gottesman1997,Freedman2001,Raussendorf2007,Fowler2012,Terhal2015}. At the heart of the procedure is a continued sequential measurement of all stabilizers $\mathcal{O}_n$. This operation serves to project the code onto a well-defined simultaneous eigenstate of the stabilizer system. As far as memory access or Clifford operations are concerned, 
occasional flips of individual stabilizers due to thermal or other fluctuations need 
not be actively corrected; they are simply \emph{recorded} by classical control.  
This simplification follows from the Gottesman-Knill theorem \cite{Gottesman1998}, 
and the reduced need for active error correction is one of the principal advantages of the 
scheme. \emph{Information} qubits are imprinted into  the code matrix either by ceasing the
measurement at individual (or groups of) physical qubits, or by creating  physical
defects through the removal of  tunneling links [cf.~the dark area in
Fig.~\ref{f1}(d)]. In either case, a `hole' is punched into the system, and the
binary eigenstates of a suitably constructed $\Bbb{Z}_2$ Wilson loop operator
\cite{Fowler2012,Wen2015} surrounding the hole then define an information qubit.
Finally, the states of both physical and information qubits must be adjustable via
the controlled flipping of individual stabilizer states. 
With these operations in place, all quantum gates required for universal quantum computation can be implemented. 
Using magic state distillation \cite{Bravyi2005}, this can be done in a fault-tolerant manner \cite{Terhal2015}.
Rather high microscopic error rates are tolerable if only the ratio in the number of physical to information qubits is
sufficiently large. It is evident, then, that the efficiency of quantum information processing crucially depends on the availability of simple stabilizer readout and
manipulation protocols, satisfying the criteria given in the Introduction. In the following, we discuss how such protocols can be implemented.

\begin{figure}[t]
  \centering
\includegraphics[width=8.5cm]{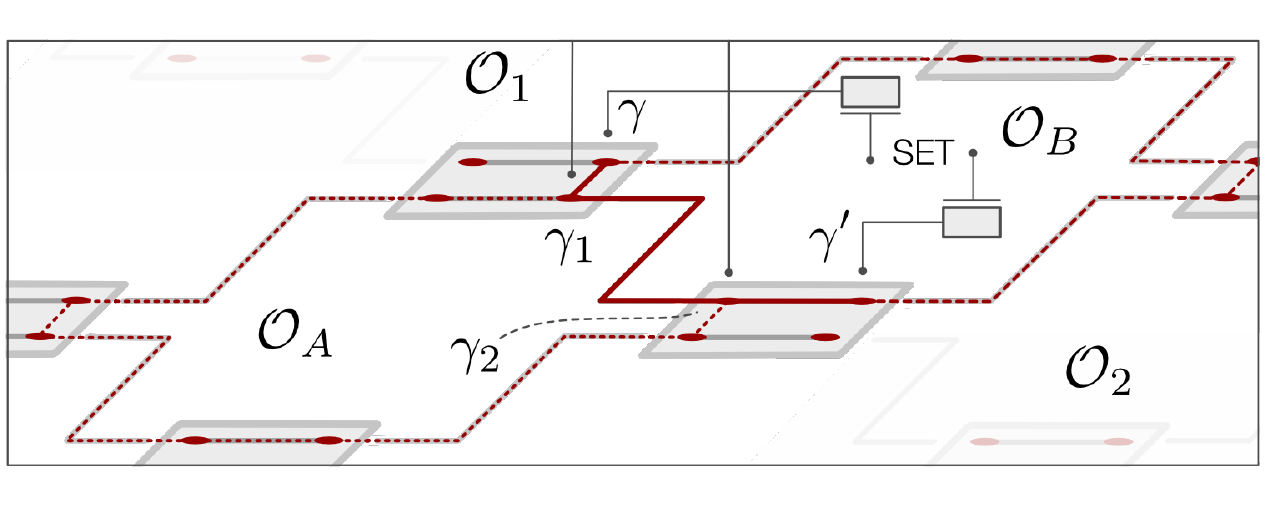}
  \caption{Surface code setup. Pairs of normal leads (indicated by vertical lines for  clarity) are  tunnel 
coupled to MBSs $\gamma_1$ and $\gamma_2$ located on different boxes. The two-terminal conductance measurement 
provides information about plaquettes ${\cal O}_A$ and ${\cal O}_B$, cf.~Eq.~\eqref{conductance}. Using SETs in the external
circuit, plaquettes can be flipped as discussed in the text. }
 \label{f2}
\end{figure}

\textit{Measurement layout.---}We consider the setup in Fig.~\ref{f2}, under the assumption that normal electrodes  are connected to individual MBSs. (For a more concrete discussion of a device architecture, see below.) A tunnel coupling of amplitude $\lambda_j$ between the Majorana fermion $\gamma_j$ and a lead electrode (with fermion operator $\Psi_j$ at the contact) is modeled by the Hamiltonian \cite{AltlandEgger2013,Beri2013},
\begin{equation}\label{htj}
H^{(j)}_t= \lambda_j \Psi_j^\dagger e^{-i\varphi_j/2}\gamma_j + {\rm h.c.},
\end{equation}
where $e^{-i\varphi_j/2}$ lowers charge by one unit on the box hosting $\gamma_j$.
Since $\gamma_j$ anticommutes with the two plaquette operators ${\cal O}_n$ containing this MBS (and commutes with all others), $H^{(j)}_t$ flips precisely those
two plaquette eigenvalues but leaves all others untouched. It is then straightforward to compute the  current flowing between two tunnel electrodes ($j=1,2$) biased by a
transport voltage. Such a current measurement will induce plaquette flips in general, with one important exception: if the leads are attached to
neighboring MBSs  located on different boxes, see Fig.~\ref{f2}, both $\gamma_1$ and $\gamma_2$ belong to the \emph{same} pair of plaquettes, ${\cal
O}_{n=A,B}$. The corresponding plaquette eigenvalues are thus flipped twice (and
hence remain invariant) whenever an electron is transported through the system. 

To make these statements quantitative, we project out all higher box charge states by Schrieffer-Wolff transformation, $H_C+H_t+ \sum_{j=1,2}H_t^{(j)}\to H_{\rm eff}$, where  
\begin{equation}\label{Heff}
H_{\rm eff}=\alpha \left( \xi + c^\ast_A {\cal O}_A+c_B{\cal O}_B\right)
\Psi_1^\dagger \Psi_2^{} + {\rm h.c.},
\end{equation}
$\alpha= - 32\lambda_1\lambda_2^*/(5 t^\ast_{12} E_C)$, and the presence of the
plaquette amplitudes $c_{n=A,B}$ of Eq.~\eqref{sysham} indicates  charge transfer
processes around either loop $A$ or $B$.  The tunneling amplitude  $\xi=[5|t_{12}|^2/(16E_C)]\eta$ describes  direct charge 
transport along the link $1\to 2$ and exceeds the loop amplitudes parametrically in $(E_C/t_{ll'})^2$. However, it is
also proportional to a `detuning parameter' $\eta$~\cite{SM}, which for small
deviation $\Delta n_{g,i}$ of the gate parameters $n_{g,i}$ off integer values scales as $\eta\sim \Delta n_{g,1} \Delta n_{g,2}$. 
Specifically, for $\Delta n_{g,i}=0$, destructive interference removes the direct amplitude $\sim\xi$ completely.  
Perturbation theory in $H_{\rm eff}$ then yields the tunnel conductance as
\begin{equation}\label{conductance}
\frac{G_{12}}{e^2/h} = 4\pi^2 |\alpha|^2 \nu_1\nu_2\left(
g_{0} + g_A {\cal O}_A + g_B {\cal O}_B +
g_{AB} {\cal O}_A {\cal O}_B \right),
\end{equation}
where $\nu_{1,2}$ are the density of states of the leads, and ${\cal O}_{A,B}$ now represent the  
eigenvalues ($\pm 1$) of the respective plaquette operators. Besides the ${\cal O}_n$-independent
contribution $g_0= \xi^2 +|c_A|^2+|c_B|^2$, the conductance contains
a number of contributions describing \emph{quantum interference} of 
direct tunneling and loop paths ($g_{A,B}$) or two-loop interference ($g_{AB}$). In a manner detailed below, 
these terms can be used to extract information on the stabilizers ${\cal O}_{A,B}$ by measuring 
the single transport coefficient $G_{12}$. 

\textit{Surface code manipulation.---}Once the code has been projected by successive stabilizer measurements to an 
eigenstate $\{\mathcal{O}_n\}$, we can manipulate it through the controlled flipping of select 
plaquettes. To this end, we  adapt a setup originally proposed for a single Majorana wire~\cite{Flensberg2011} and assume that individual 
MBSs $\gamma$ of the code are also tunnel-coupled to SETs, i.e., electronic islands sufficiently small that the occupancy of a single 
level can be controlled by a nearby gate, see Fig.~\ref{f2}.  Describing the SET-code coupling by a Hamiltonian as in
 Eq.~\eqref{htj} but with $\lambda_j\to \lambda$ and $\Psi_j(0)\to d$,  where $\lambda$ is the coupling amplitude and $d$ the 
charge annihilation operator on the SET, we consider a situation where the charge occupancy of
the SET is slowly lowered from $1$ to $0$, while that of a SET coupled to another MBS $\gamma'$ is raised from $0$ to $1$. 
This operation amounts to an adiabatic charge pumping process during which a single electron enters the code through $\gamma$ and exits through $\gamma'$.  (The adiabaticity relies on  the pumping rate being slower
than the inverse of an effective inter-SET coupling amplitude specified below.) The intra-code charge transfer is described by a 
linear superposition of `string operators' $\hat S$, i.e.,  products of tunneling amplitudes $\gamma_l t^\ast_{ll'}\gamma_{l'}$ 
establishing a path connection between the boxes hosting $\gamma$ and $\gamma'$, respectively. For example, in the situation shown in Fig.~\ref{f2}, the shortest string operator is given by $\hat S= \gamma_1 (2t_{12}^*/E^2_C)\gamma_2$. Each individual tunneling along the path amounts to a
virtual excitation of the system, such that a string operator involving $n$ steps scales $\sim t^n/E_C^{n+1}$ --- the process is dominated by the shortest path connecting the boxes. 
However, more important in the present context is that all string operators commute with the stabilizer system, 
$[\hat S,\mathcal{O}_n]=0$, which in turn implies that the action of the full tunneling operator, $\hat X= \lambda^\ast\lambda' \ d
\gamma \left( \sum \hat S\right) \gamma' d^{\prime \dagger}$, on the code eigenstate is effectively described by
the pair operator $\gamma\gamma'$. Now each MBS is an element of two diagonally
adjacent plaquettes, which implies that, depending on the separation between $\gamma$ and $\gamma'$, a 
pair $(\gamma,\gamma')$  has either two, one, or no plaquettes in common \cite{foot2}. 
In the case of one shared plaquette, exemplified in Fig.~\ref{f2}, the SET operation will create a minimal two stabilizer excitation ($\mathcal{O}_{1,2}$ in Fig.~\ref{f2}). For larger  separation between the excitation centers $\gamma,\gamma'$, four stabilizers get flipped. (Increasing the separation also diminishes the effective inter-SET coupling 
described by $\hat X$, and this imposes tighter conditions on the adiabaticity of the process \cite{footnew}.) Finally, successive SET pumping operations can be employed to \emph{move} excitations.

\textit{Operation of the system.---} The first step in the operation of the system containing $N$ plaquettes must be a  one-time calibration operation 
in which the non-universal but fixed transport coefficients $g_{0,A/B,AB}$ appearing in the  conductance Eq.~\eqref{conductance} are determined for all links 
of the system.  This step involves the measurement of the corresponding tunnel conductances  followed by the flipping of one of the neighboring plaquettes \cite{SM}.  
Once these coefficients are known, the projective \cite{foot3} measurement of $O(N)$ suitably chosen tunnel conductances provides the full 
information on the stabilizer system $\{{\cal O}_n\}$. Active operations can then be performed by SET-induced changes of stabilizers, by stopping the measurement of certain stabilizers, or by the physical removal of bonds, cf.~Ref.~\cite{Fowler2012}.

\begin{figure}[t]
  \centering
\includegraphics[width=8.5cm]{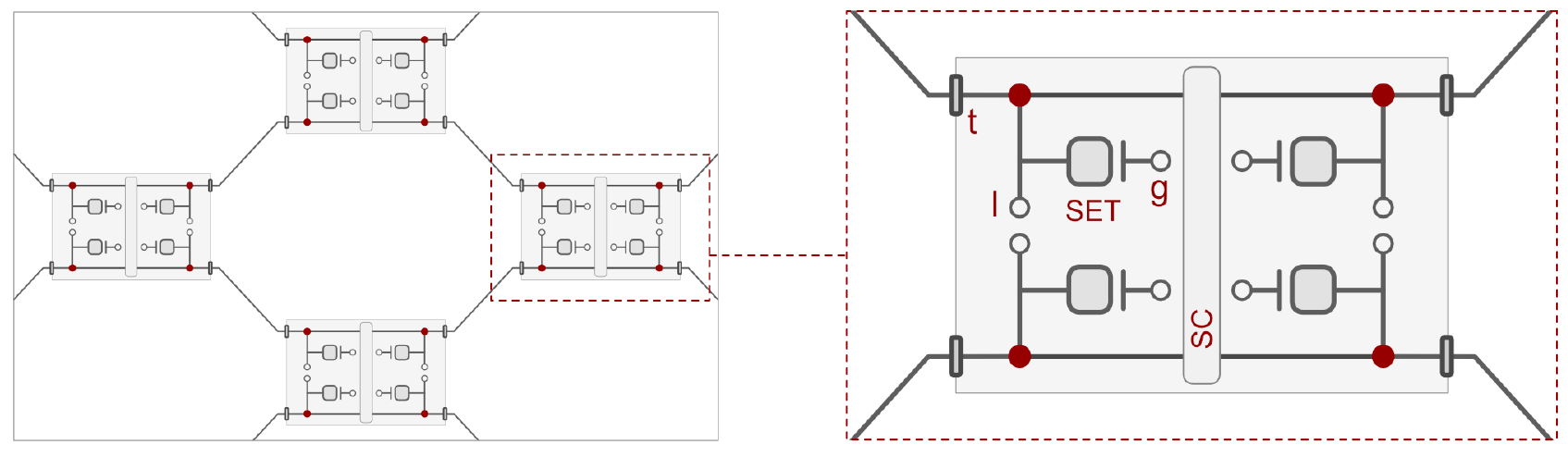}
  \caption{Concrete design of a 2D on-chip implementation. The Majorana-Cooper box hosts four MBSs (red dots) and is implemented as 90~deg rotated H-structure from a pair of second-generation topological nanowires \cite{Chang2015} joined by a mesoscopic superconducting slab (`SC').  This slab ensures a common charging energy for the box. 
One normal lead (`$|$') and one SET (along with its gate electrode, `g') are tunnel-coupled  to each MBS by employing T-junctions (see text), such that there are eight contacts per box. MBSs on 
neighboring boxes are connected by short wires via tunnel contacts (`t'). \label{f3}}
\end{figure}

\textit{Towards an operational hardware.---}Finally, let us suggest a hardware blueprint closer to current-date experimental reality than the schematic drawings in Figs.~\ref{f1} and \ref{f2}. The key ingredient of the setup, single MBSs tunnel-coupled to neighboring MBSs and at the same time to a normal lead and a SET, could be implemented by so-called Majorana T-junctions \cite{CMpriv}, 
cf.~the junction geometry sketched in Fig.~\ref{f3}. Such T-junctions are instrumental to Majorana braiding protocols \cite{Alicea2012,Leijnse2012,Beenakker2013} and a concrete roadmap for their realization via crossed nanowires exists~\cite{CMpriv}.  With the aid of T-junctions, the contact to an external probe and the SET associated to the MBS would then be implemented nearby and within the 2D structure. A four-fold replication of this unit allows contact to the four MBSs on a Majorana-Cooper box, realized by the small capacitive H-structure in Fig.~\ref{f3}. In a next step, one might isolate the entire array of tunnel-coupled boxes via a 2D oxide layer, and connect the external (lead and SET gate) contacts of each box to electrodes approaching from above. The implementability of a minimal qubit based
on this idea could be tested on a two-loop prototype with six connected boxes, cf.~the layout in Fig.~\ref{f2}, which allows one to change the system state by SET
operations and to measure the ensuing change in the tunnel conductance. (With minimal changes in the design, one might  reduce the number of connected boxes to three.)
Finally, the so-called quasiparticle poisoning time limiting the coherence of MBSs is found to be of order 10~ms \cite{Higginbotham2015}. This timescale may set a present-date limit to the duration of a full operation cycle, i.e., to the measurement of all stabilizers.

\textit{Concluding remarks.---}In this Letter, we have suggested a hardware layout and concrete protocols for the stabilization and manipulation of Majorana surface codes. 
Salient features of the proposal include implementability entirely in terms of device technology that exists already or is subject to current experimental effort, a single-step measurement protocol based on tunneling conductance measurements, and independence of microwave radiation and/or magnetic interferometry.  Given the current pace of progress, we are optimistic that a prototypical qubit based on these ideas might be realizable soon. From there, it will still be a long shot to a large scale device.  However, the absence of external electromagnetic fields in the operation of the system implies a level of protective isolation that can ultimately provide a powerful aid towards scalability.

\textit{Acknowledgments.---}We thank J. Eisert, F. Hassler, K. Flensberg, L. Fu and C. Marcus for discussions, and acknowledge funding by the Israel Science Foundation Grant No.~1243/13 (E.S.), by the Marie Curie CIG Grant No.~618188 (E.S.),  by Microsoft Project Q (S.M.A.), and by the Deutsche Forschungsgemeinschaft Grant No.~EG~96/10-1 within network SPP 1666 (R.E.).

\clearpage
\setcounter{equation}{0}
\renewcommand{\theequation}{S.\arabic{equation}}
\setcounter{figure}{0}
\renewcommand{\thefigure}{S.\arabic{figure}}

\begin{widetext}
\begin{center}
\section*{Towards realistic implementations of a Majorana surface code - Supplementary Material}
In this supplementary material we include a detailed derivation of the effective tunnelling Hamiltonian and conductance for the noninvasive measurement.
\end{center}
\end{widetext}

We here sketch the derivation of the effective Hamiltonian [Eq.~(3) in the Letter] and the resulting conductance [Eq.~(4)]
using the perturbative Schrieffer-Wolff (SW) procedure (see Ref.~[24] of the Letter for an introduction).  
For brevity, let us consider the example of tunneling via the 
`direct link' $\gamma_1\gamma_2$ between leads $1$ and $2$, see Fig.~2
in the Letter. The relevant parts of the Hamiltonian are denoted as
\begin{eqnarray}
H_t &=& -\frac12 t_{12}\gamma_1\gamma_2 
e^{i(\varphi_1-\varphi_2)/2} ~+\mathrm{h.c.}~,\notag\\\label{eq1}
H_{t}^{(j=1,2)} &=& \lambda_j\Psi^\dagger_j \gamma_j 
e^{-i\varphi_j/2}~+\mathrm{h.c.}~,\\\notag
H_C &=& \sum_{j=1,2} E_{C,j} \left(\mathcal{N}_j -n_{g,j}\right)^2,
\end{eqnarray}
where $\mathcal{N}_j$ and $\varphi_j$ are the charge number 
operators on the two boxes and their conjugate phases, respectively.
Now consider backgates which are set close to integer charges $Q_j$, 
i.e. $n_{g,j}=Q_j+\Delta n_{g,j}$ with $|\Delta n_{g,j}|\ll 1$. 
The energy needed to add (remove) a single charge to the $j$th island is then 
given by $E_j^{\pm} = (1\mp 2\Delta n_{g,j})E_{C,j}$.
We may now restrict ourselves to sequences of tunneling which 
implement an effective transfer from lead $2$ to lead $1$, i.e.,
$\sim\Psi_1^\dagger\Psi_2$, since the other direction follows by 
Hermitian conjugation. The minimal process has to involve each of the 
operators $L_1 \sim \Psi_1^\dagger\gamma_1e^{-i\varphi_1/2}$, 
$L_2^\dagger \sim \gamma_2\Psi_2e^{i\varphi_2/2}$, 
and $A_{12} \sim \gamma_1\gamma_2 e^{i(\varphi_1-\varphi_2)/2}$ once. As an example, the sequence $L_1A_{12}L_2^\dagger$ first brings us to an excited state of energy $E_2^+$ by adding an electron to the box hosting $\gamma_2$; subsequently, the electron is transferred to the box of $\gamma_1$, yielding the scale $E_1^+$, last it is removed via lead $1$. The relevant energy denominator hence follows as $1/(E_1^+E_2^+)$.
Taking all six possible permutations, corresponding to 
different time ordering of tunneling events, and summing up 
the excitation energy denominators, it is convenient to define 
the `detuning parameter' $\eta$ used in the Letter,
\begin{eqnarray}
\frac{(E_1^+-E_1^-)(E_2^+-E_2^-)}{E_1^+E_1^-E_2^+E_2^-}&\\\notag
= \frac{4}{E_{C,1} E_{C,2}}
\frac{4\Delta n_{g,1}\Delta n_{g,2}}{(1-4\Delta n_{g,1}^2)(1-4\Delta n_{g,2}^2)} &&=: \frac{4}{E_{C,1} E_{C,2}}\eta
\end{eqnarray}
The effective tunneling amplitude mediated by the `direct link' 
$\sim\gamma_1\gamma_2$, with the
tunneling amplitudes in Eq.~\eqref{eq1}, then follows as
\begin{equation}
H_{12,\mathrm{eff}} = \tau \Psi^\dagger_1\Psi_2 ~+\mathrm{h.c.},\quad
\tau = -\frac{2\lambda_1\lambda_2^\ast t_{12}}{E_{C,1}E_{C,2}} \eta.
\end{equation}
We observe that in order to obtain a finite tunneling amplitude 
$\tau$, we have to detune both backgates $n_{g,j=1,2}$ away from 
integer values, while $E_{C,1}\neq E_{C,2}$ has no substantial effect.
(We thus assume isotropic $E_C$ below and in the main text.)
Following the same steps for tunneling around the 
plaquettes $\mathcal{O}_{A,B}$ containing the pair 
$\gamma_1\gamma_2$, cf.~Fig.~2, we obtain the full effective Hamiltonian as 
stated in Eq.~(3) of the Letter.
Remarkably, the parameter $\xi = \tau/\alpha$ used in that equation
contains no complex phase and can be changed through the detuning
parameter $\eta$. Note that we have neglected the detuning effect 
on the plaquette contributions, i.e., on the terms $\sim c_n$ appearing in 
Eqs.~(1) and (3) of the Letter. While this can be easily included, it 
has no significant impact on the physics nor on the operation of the code.
Calculating the tunnel conductance by perturbation theory in $H_{\rm eff}$, 
we obtain Eq.~(4) in the Letter.

A measurement of the conductance $G_{12}$ hence contains terms 
proportional to $\mathcal{O}_A$, $\mathcal{O}_B$ 
(with coefficient $g_{A,B} =$\\$2\xi \mathrm{Re}(c_{A,B})$) and $\mathcal{O}_{A}\mathcal{O}_{B}$ (with coeff. $g_{AB} = 2\mathrm{Re}(c_Ac_B)$). 
Using a sequence of conductance measurements $G_{12}$ and controlled 
plaquette flips on the A or B plaquette, each of these values may be 
extracted separately: the sum of two measurements 
$G_{12}(\mathcal{O}_B)$ and $G_{12}(-\mathcal{O}_B)$ 
reveals $\mathcal{O}_A$. Additionally switching between 
$\mathcal{O}_A$ and $-\mathcal{O}_A$ determines the 
corresponding conductance coefficient $g_A$.\\
For good visibility of conductance switches, we may set 
$|\xi| \approx |c_{A,B}|$ by adjusting the 
detuning $\eta$. Assuming we are deep in the charge quantization
regime, $|t|/E_C \ll 1$, this results in 
the condition $\Delta n_{g,1}\Delta n_{g,2} \approx |t|^2/(4E_C^2)$. 
For some $|t|/E_C \lesssim 1/3$ yielding reasonable plaquette energies, the necessary detuning
away from the Coulomb valley center can hence be estimated 
as $|\Delta n_{g,j}| \lesssim 0.16$, still sufficiently far from half-integer.


\begin{thebibliography}{99}
\bibitem{foot1}
Random but static sign changes of $\mathrm{Re}(c_n)$ can be absorbed by a sign change of $\mathcal{O}_n$. Similarly, $E_C$ fluctuations between different boxes do not spoil functionality \cite{SM}.
\bibitem{foot2}
For stabilizer readout, with $\gamma=\gamma_1$ and $\gamma'=\gamma_2$ in Fig.~\ref{f2}, there are three string operators $\hat S\sim 1,{\cal O}_A,{\cal O}_B$, whose interference yields Eq.~\eqref{conductance}.
\end{thebibliography}
\end{document}